\documentclass[prb,twocolumn]{revtex4}
\usepackage{graphics}
\usepackage{picinpar}
\usepackage{epsfig}
\usepackage{bbm}
\usepackage{float}
\usepackage{amsmath}

\begin{document}
\title{The $m$-component spin glass on a Bethe lattice}
\author{A. Braun}
\author{T. Aspelmeier}
\affiliation{Institut f\"ur Theoretische Physik, Universit\"at G\"ottingen, Friedrich-Hund-Platz 1, 37077 G\"ottingen}

\begin{abstract}
We study the $m$-component vector spin glass in the limit $m\to\infty$ on a Bethe lattice. The cavity method allows for a solution of the model in a self-consistent field approximation and for a perturbative solution of the full problem near the phase transition. The low temperature phase of the model is analyzed numerically and a generalized Bose-Einstein condensation is found, as in the fully connected model. Scaling relations between four distinct zero-temperature exponents are found.
\end{abstract}

\maketitle

\section{Introduction}
The spin glass with $m$-component spins in the limit of large $m$ is an interesting member of the spin glass family for various reasons. First, in the limit $N\to\infty$ it is replica symmetric\cite{Almeida:1978b}. This sets it apart from high-dimensional and mean-field spin glasses with Ising, XY or Heisenberg spins but makes it more related to low-dimensional spin glasses, provided the replica symmetric droplet picture holds in low dimensions. Second, it is analytically more tractable than, say, Ising systems since the limit $m\to\infty$ allows for some simplifications, and since the replica symmetric theory is much simpler than the theory for broken replica symmetry. Third, it has been shown\cite{Hastings:2000,Aspelmeier:2004a,Lee:2005a} to have an unusual type of phase transition, namely a generalized Bose-Einstein condensation where the spins condense in a high dimensional subspace in the low temperature phase. Finally, it has been studied in a field-theoretic approach in high dimensions\cite{Viana:1988,Green:1982,Viana:1993}, and by computer simulations in low dimensions\cite{Morris:1986,Lee:2005b}.

In this work we study the $m$-component spin glass on a Bethe lattice. The Bethe lattice allows for some exact analytical results using the cavity method\cite{Mezard:2001} and the influence of the finite connectivity can be analyzed explicitly. The cavity method has to be extended since cavity fields alone are not sufficient to describe the behaviour of such a spin system and terms of quadratic and higher order need to be taken into account. With this modification, we find a phase transition to a replica symmetric state in the model at a finite temperature.

The generalized Bose-Einstein condensation is also observed on the Bethe lattice. It is characterized by an exponent $\mu$ which describes the scaling of the dimensionality $n_0$ of the ground state subspace with the number of spins $N$. For the fully connected model $\mu=2/5$\cite{Hastings:2000,Aspelmeier:2004a}; here we find different and connectivity-dependent values. In addition to $\mu$ there are a number of other zero temperature exponents: an exponent we call $x$ describes the scaling of the ground state energy $e(m,N)$ per spin and component with $m$ for $N=\infty$, the scaling of $e(m,N)$ with $N$ for $m=\infty$ is characterized by an exponent $y$ and $\nu$ is the exponent of the singular part of the eigenvalue spectrum of the inverse susceptibility matrix (for $m=N=\infty$). In this paper we show that these four exponents are not independent but related by scaling laws. As noted in Ref.\ \onlinecite{Lee:2005a}, the order of limits $m\to\infty$ and $N\to\infty$ is important. Physically, taking $N\to\infty$ first makes most sense. Analytically, it is often more useful to take the opposite order, and this is what we will do in this work. Moreover, the generalized Bose-Einstein condensation can only be observed for the order $m\to\infty$ first. The scaling relations show that even though the order of limits is important, it is possible to obtain information from one order of limits about the other. 

The paper is organized as follows. In Sec.~\ref{model} we present the details of our model. In Sec.~\ref{cavity} we extend the cavity method to $m$-component spins, which we need in Sec.~\ref{phase} to study the phase transition. In Sec.~\ref{gbec} we examine the generalized Bose-Einstein condensation numerically. The results of this are used to check the scaling relations derived in Sec.~\ref{scaling}. We conclude in Sec.~\ref{conclusion}. 

\section{Model}
\label{model}
The spin glass model we analyse in this work consists of $m$-component vector spins $\vec s_i$ on a Bethe lattice. The spins have fixed length $|\vec s_i|^2=m$. The Bethe lattice in the context of disordered systems is a random graph with a fixed connectivity for every spin, equal to $k+1$ in this work. In such a graph consisting of $N$ sites, loops are of order $\log N$. Therefore the structure of the lattice is locally tree like. We analyse the model in the $m\rightarrow\infty$-limit for which the fully connected spin glass is replica symmetric\cite{Almeida:1978b}. The same is expected to hold on the Bethe lattice and we will restrict our work to that case. The generic Hamiltonian for the described model is
\begin{eqnarray}
H=-\sum_{ij}J_{ij}\vec s_i\vec s_j.
\end{eqnarray}
Here the exchange interactions $J_{ij}$ are equal to zero if the
spins $\vec s_i$ and $\vec s_j$ are not nearest neighbours on the
Bethe lattice. If they are, $J_{ij}$ is drawn from a distribution $P(J_{ij})$ which is either a bimodal distribution (corresponding to the $\pm J$ model) or a Gaussian distribution. The width of these distributions $J'$ will be chosen to scale with the number of neighbours, such that $J'=J/\sqrt{k}$ with $J$ fixed. This ensures easy comparison in the limit $k\to\infty$ with the fully-connected spin glass where $J'=J/\sqrt N$.

\section{Cavity method for $m$-component spins}
\label{cavity}

\begin{figure}
\includegraphics[width=\linewidth]{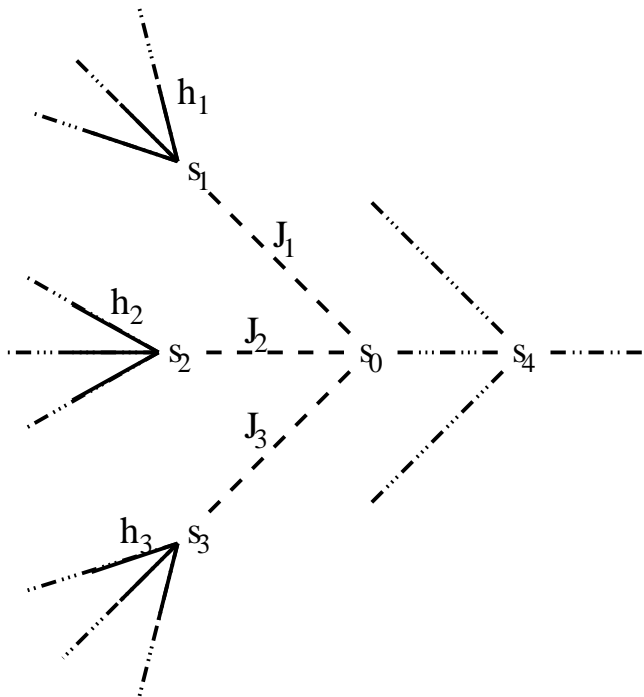}
\caption{The three spins $\vec s_1$, $\vec s_2$ and $\vec s_3$ and their back branches are merged onto spin $\vec s_0$ with coupling constants $J_1$, $J_2$ and $J_3$ in this example with $k=3$. In the next step of the iteration, $\vec s_0$ and two other branches will be merged onto the spin $\vec s_4$.}
\label{merge}
\end{figure}

The cavity method on a $k+1$-connected Bethe lattice for Ising spins is explained in detail in Ref.\ \onlinecite{Mezard:2001} and we refer the reader to this paper for a review. The main idea for Ising spins is that the local field at a site in the absence of one of its neighbours (the cavity field) is a quantity which can be propagated through the lattice iteratively. This is done by merging $k$ branches of the lattice onto a new spin $s_0$, see Fig.~\ref{merge}. The many back spins in each branch may be represented by an effective cavity field acting on the end spins $s_i$. Before merging, the end spins $s_i$ ($i=1,\dots,k$) of these branches are each missing a neighbour, but after the merger they are $k+1$-connected. The new spin $s_0$ only has $k$ neighbours from the $k$ branches that are being joined and is thus of the same type as the merged spins $s_1,\dots,s_k$ were before the merger.  The original spins $s_1,\dots,s_k$ are then traced out, leaving only $s_0$ with a new effective cavity field (which of course depends on the original $k$ cavity fields). This procedure may then be iterated, leading to a recursion relation for the cavity fields which can be used to calculate their distribution $Q(\vec h)$, see Sec.\ \ref{phase}.

In the case of $m$-component spins this method needs to be modified because a simple vector field is not sufficient to describe the action of $k$ old spins and their effective fields on a new spin. We will see below why this is the case. For the time being, we write the effective Hamiltonian $H_i$ of a $k$-connected spin $i$ in general form as
\begin{align}
H_i &= -\vec h_i \vec s_i - \vec {s_i}^{\text T} A_i \vec s_i - B_i(\vec s_i),
\label{genham}
\end{align}
i.e.\ a linear term with a field $\vec h_i$, a quadratic term with a symmetric traceless matrix $A_i$, and the rest, contained in the function $B_i(\vec s)$ which has only terms of order $s^3$ or higher. The matrix $A_i$ may be chosen to be traceless without loss of generality because any nonzero trace only yields a constant contribution due to the constraint $\vec {s_i}^2=m$. When $k$ such spins are merged onto a new spin $s_0$ with coupling constants $J_i$, the partition function of this system is
\begin{align}
Z_0 &= \int \left(\prod_{i=0}^k d^m s_i\,\delta(m-\vec {s_i}^2)\right) \exp\left\{-\beta \sum_{i=1}^k (-J_i \vec s_0 \vec s_i + H_i) \right\} \\
&= \int d^m s_0\,\delta(m-\vec {s_0}^2) \prod_{i=1}^k Z_i, \text{ where} \label{Z0}\\
Z_i &= \int d^m s_i\,\delta(m-\vec {s_i}^2) \exp\left\{ -\beta H_i\big|_{\vec h_i\to \vec h_i+J_i\vec s_0} \right\}.
\end{align}
Here $\beta=1/k_B T$ is the inverse temperature, as usual.
The task is to calculate $Z_i$ and express $\prod_{i=1}^k Z_i$ in the
form $\exp\{-\beta H_0\}$ where $H_0$ has the same functional form as
in Eq.~\eqref{genham}. 
We will now proceed in two ways. The first is a self-consistent effective field approach where we use the description solely in terms of fields regardless of the fact that it is not sufficient as mentioned above. The effective field will be determined by the condition that it has to reproduce the spin expectation value of the full Hamiltonian. This procedure will show that there is a spin glass transition at a finite temperature. The second way is to analyze the full problem, which is only possible in the vicinity of the transition. When that is done, it will become clear that near the transition the self-consistent field approach is equivalent to the first order approximation of the full problem. It may thus be viewed as the first step of a systematic approximation for the whole low temperature phase.

\subsection{Self-consistent effective field approach}

Setting $A_i=B_i(\vec s)=0$, we have to calculate 
\begin{align}
Z_i &= \int d^m s_i\,\delta(m-\vec {s_i}^2)\exp\left\{\beta (\vec h_i +J_i \vec s_0) \vec s_i \right\}.
\label{Zi0}
\end{align}
For notational simplicity, we will abbreviate the term $\vec h_i+J_i \vec s_0$ by $\tilde{\vec h}_i$. We then have
\begin{align}
Z_i&= \int_{-i\infty+c}^{i\infty+c} \frac{dz}{2\pi}\int d^m s_i\,\exp\left\{\beta \tilde{\vec h}_i \vec s_i + z(m-\vec {s_i}^2)\right\}
\label{Zi1} \\
&= \int_{-i\infty+c}^{i\infty+c} \frac{dz}{2\pi}\exp\left\{\beta^2\tilde{\vec h}_i^2/4z+zm-\frac m2 \ln \frac{z}{\pi}\right\}.
\label{Zi}
\end{align}
Here we have used an integral representation of the $\delta$-function
and $c$ is an arbitrary positive constant to ensure convergence of the
Gaussian integrals over the spin components. The remaining integral
over $z$ may be evaluated in the limit $m\to\infty$ using steepest
descent methods. Introducing $\tilde{\vec h}_i'=\tilde{\vec h}_i/\sqrt
m$ and $\vec {s_0}'=\vec s_0/\sqrt m$, the exponent in the integral in
Eq.~\eqref{Zi} is $m(\beta^2(\tilde{\vec h}_i')^2/4z+z-\frac 12\ln
z)$, apart from irrelevant constants. 
The saddle point equation is therefore
\begin{align}
\frac{\partial}{\partial z} (\beta^2(\tilde{\vec h}_i')^2/4z+z-\frac 12\ln z) &= 1-\beta^2(\tilde{\vec h}_i')^2/4z^2 - \frac{1}{2z} = 0.
\end{align}
The relevant solution of the saddle point equation is $ z = z_i^0 =
\frac 14\left(1+\sqrt{1+4\beta^2(\tilde{\vec h}_i')^2}\right)$ and we get
\begin{align}
Z_i &= \exp\left\{m(\beta^2(\tilde{\vec h}_i')^2/4z_i^0+z_i^0-\frac 12\ln z_i^0) \right\}.
\label{Zi2}
\end{align}
Note that according to Eq.~\eqref{Zi1} this is nothing but the partition function of a single spin in a field $\tilde{\vec h}_i$. Therefore one can obtain the spin expectation value in a field by taking the derivative with respect to $\tilde{\vec h}_i$,
\begin{align}
\langle \vec s_i\rangle &= \frac{1}{\beta}\frac{\partial \ln Z_i}{\partial \tilde{\vec h}_i} = \frac{\beta \tilde{\vec h}_i}{2z_i^0} = \frac{2\beta \tilde{\vec h}_i}{1+\sqrt{1+4\beta^2(\tilde{\vec h}_i')^2}}.
\label{spinavg}
\end{align}
We will need this result below.

We can now calculate the propagated partition function $Z_0$. For that purpose, the term $J_i\vec s_0$ that was hidden in $\tilde{\vec h}_i$ must be reintroduced, i.e.\ we have $z_i^0=\frac 14\left(1+\sqrt{1+4\beta^2(\vec h_i'+J_i\vec {s_0}')^2}\right)$ in the following. According to Eqs.~\eqref{Z0} and \eqref{Zi2}, the partition function for spin $\vec s_0$ is
\begin{widetext}
	\begin{align}
	Z_0 &= \int d^m s_0\,\delta(m-\vec {s_0}^2)\exp\left\{ m\sum_{i=1}^k(\beta^2(\vec h_i'+J_i\vec {s_0}')^2/4z_i^0+z_i^0-\frac 12\ln z_i^0)\right\}.
	\label{Z01}
	\end{align}
\end{widetext}
Here it can be seen explicitly that $Z_0$ is not the partition function of a single spin in a field and that the cavity method does not close on the level of fields. The self-consistent approximation now consists of finding a field $\vec h_{0}$ which generates the same spin expectation value as the partition function $Z_0$. In order to calculate the latter (the former has already been calculated above in Eq.~\eqref{spinavg}), we again need to employ steepest descent methods, but this time not only for the auxiliary variable $z$ from the integral representation of the $\delta$-function but also for all $m$ spin components $s_0^\alpha$. Here the problem arises that the number of integration variables, $m+1$, is of the same order as the large parameter in the integrand, $m$. In such a situation steepest descents can not be used. The situation is remedied, however, by the following observation. The spin $\vec s_0$ only appears in the combination $(\vec h_i+J_i\vec s_0)^2=\vec h_i^2 + 2J_i\vec h_i\vec s_0 + mJ_i^2$, so all components of $\vec s_0$ orthogonal to the $\vec h_i$ are projected out. These $m-k$ orthogonal components may be integrated out first as they are merely Gaussian integrals, leaving only $k+1$ nontrivial integrals which may then be treated with the saddle point method. Denoting by $O$ the orthogonal transformation $\vec t = O\vec s_0$ which transforms the spin variables such that only the integrals over the first $k$ components remain while the rest can be carried out (the details of this transformation are unimportant, as will become apparent below), we get $(\vec h_i+J_i\vec s_0)^2=\vec h_i^2+2J_i(O\vec h_i)\vec t + mJ_i^2$ and $z_i^0 = \frac 14\left(1+\sqrt{1+4\beta^2((\vec h_i')^2+2J_i(O\vec h_i'){\vec t}\hspace{1ex}' + J_i^2)}\right)$ with ${\vec t}\hspace{1ex}'=\vec t/\sqrt m$. The vector $\vec t$ initially has $m$ components, but after carrying out the integrals over the last $m-k$ components only $k$ components remain and one gets
\begin{widetext}
	\begin{align}
	Z_0 &= \int_{-i\infty}^{i\infty}\frac{dz}{2\pi} \int d^k t\,\exp\left\{ m\sum_{i=1}^k(\beta^2((\vec h_i')^2+2J_i(O\vec h_i'){\vec t}\hspace{1ex}' + J_i^2)/4z_i^0+z_i^0-\frac 12\ln z_i^0+z(1-({\vec t}\hspace{1ex}')^2))-\frac{m-k}{2}\ln\frac{z}{\pi}\right\}.
	\label{Z02}
	\end{align}
\end{widetext}
The saddle point equations following from this are 
\begin{align}
z &= \frac{1}{2(1-({\vec t}\hspace{1ex}')^2)}\label{saddle1} \\
2z\vec t &= \sum_{i=1}^k\frac{2\beta^2 J_i O\vec h_i}{4z_i^0}.
\label{saddle2}
\end{align}
Since $\vec t$ is merely the transformed spin $\vec s_0$, the saddle point value of $\vec t$ immediately gives the expectation value of the spin via
$\langle \vec s_0 \rangle = O^T\vec t$ (padding the last $m-k$ components of $\vec t$ with zeroes). Multiplying Eq.~\eqref{saddle1} by $2\langle \vec s_0\rangle$ yields
\begin{align}
2z\langle \vec s_0\rangle &= \frac{\langle \vec s_0\rangle}{1-\langle \vec {s_0}'\rangle^2},
\label{saddle1b}
\end{align}
while applying $O^T$ from the left to Eq.~\eqref{saddle2} results in
\begin{align}
	2z\langle \vec s_0\rangle &= \sum_{i=1}^k\frac{\beta^2 J_i \vec h_i}{2z_i^0}.
	\label{saddle2b}
\end{align}
Inside $z_i^0$, $\vec t$ may likewise be replaced by  $O\langle\vec s_0\rangle$, such that $z_i^0 = \frac 14\left(1+\sqrt{1+4\beta^2((\vec h_i')^2+2J_i\vec h_i'\langle \vec {s_0}'\rangle + J_i^2)}\right)$.

We are now in a position to present the equation which determines the effective field. It follows from Eq.~\eqref{spinavg} that, if the spin $\vec s_0$ were only subject to a field $\vec h_{0}$, one would have $\beta \vec h_{0} = \frac{\langle \vec s_0\rangle}{1-\langle\vec {s_0}'\rangle^2}$. Combining this with Eqs.~\eqref{saddle1b}, \eqref{saddle2b} and again Eq.~\eqref{spinavg} one gets
\begin{align}
\vec h_{0} &= \sum_{i=1}^k\frac{2\beta J_i \vec h_i}{1+\sqrt{1+4\beta^2((\vec h_i')^2+4\beta J_i\frac{\vec h_i' \vec h_{0}'}{1+\sqrt{1+4\beta^2(h_{0}')^2}}+J_i^2)}},
\label{h0}
\end{align}
with $\vec h_{0}'=\vec h_{0}/\sqrt m$. This complicated implicit equation determines the field $\vec h_{0}$ which produces the same spin expectation value as $Z_0$. It corresponds to the Ising analogue $h_0 = \sum_{i=1}^k u(J_i,h_i)$ from Ref.\ \onlinecite{Mezard:2001} where $u(J_i,h_i)=\frac{1}{\beta}\,\mathrm{atanh}(\tanh \beta J_i \tanh \beta h_i)$. In our context, the presence of $\vec h_0$ on the right hand side of Eq.~\eqref{h0} makes the solution much more difficult. Above and close to the phase transition, however, an expansion of Eq.~\eqref{h0} in powers of $\vec h_i$ should be valid. To first order this yields
\begin{align}
\vec h_0 &= \sum_{i=1}^k\frac{2\beta J_i \vec h_i}{1+\sqrt{1+4\beta^2 J_i^2}}.
\label{h01}
\end{align}

\subsection{Analysis of the full problem}

We now turn to the full problem with the Hamiltonian from Eq.~\eqref{genham}. As before for the self-consistent effective field approach, we will denote the term $\vec h_i+J_i\vec s_0$ by $\tilde{\vec h}_i$ in the partition function $Z_i$. The partition function $Z_i$ may be calculated perturbatively in the following way:
\begin{widetext}
	\begin{align}
	Z_i &= \int_{-i\infty+c}^{i\infty+c}\frac{dz}{2\pi}\int d^m s_i\, \exp\left\{ \beta(\tilde{\vec h}_i \vec s_i + \vec {s_i}^T A_i \vec s_i + B_i(\vec s_i)) + z(m-\vec {s_i}^2)\right\} \\
	&= \int_{-i\infty+c}^{i\infty+c}\frac{dz}{2\pi}\int d^m s_i\, \exp\left\{ \beta(\tilde{\vec h}_i \vec s_i - \vec {s_i}^T (\frac{z}{\beta}-A_i) \vec s_i) + mz \right\}\sum_{n=0}^\infty \frac{(B_i(\vec s_i))^n}{n!} \\
	&= \int_{-i\infty+c}^{i\infty+c}\frac{dz}{2\pi} \exp\left\{mz + \frac{\beta}{4} \tilde{\vec h}_i^T \left(\frac{z}{\beta}-A_i\right)^{-1}\tilde{\vec h}_i - \frac 12 \ln\det\left(\frac{z}{\beta}-A_i\right) + \frac m2 \ln\pi \right\} \sum_{n=0}^\infty \frac{\langle B_i^n\rangle_{i}}{n!} \\
	&= \int_{-i\infty+c}^{i\infty+c}\frac{dz}{2\pi} \exp\left\{mz + \frac{\beta}{4} \tilde{\vec h}_i^T \left(\frac{z}{\beta}-A_i\right)^{-1}\tilde{\vec h}_i - \frac 12 \ln\det\left(\frac{z}{\beta}-A_i\right) + \frac m2 \ln\pi\right\} \\ &\quad\times\exp\left\{ \langle B_i\rangle_i + \frac 12 (\langle B_i^2\rangle_i-\langle B_i\rangle_i^2) + \cdots\right\}
	\label{Zi3}
	\end{align}
\end{widetext}
The angular brackets $\langle \cdot\rangle_{i}$ denote the average with respect to the weight $\exp\{\beta(\tilde{\vec h}_i \vec s_i - \vec {s_i}^T (\frac{z}{\beta}-A_i) \vec s_i)\}$. The expression $\frac{z}{\beta}-A_i$ should be interpreted as a matrix expression, i.e.\ as $\frac{z}{\beta}\mathbbm{1}-A_i$. The dots in the last line above indicate higher order cumulants. Now $Z_i$ from Eq.~\eqref{Zi3} can again be treated by steepest descent methods in the limit $m\to\infty$, and the resulting saddle point equation is
\begin{widetext}
	\begin{align}
	1 &= \frac{1}{4} (\tilde{\vec h}_i')^T \left(\frac{z}{\beta}-A_i\right)^{-2}\tilde{\vec h}_i' + \frac{1}{2m\beta}\mathrm{Tr}\left(\frac{z}{\beta}-A_i\right)^{-1} - \frac 1m \frac{\partial\langle B_i\rangle_i}{\partial z} + \cdots.
	\label{saddle5}
	\end{align}
\end{widetext}
The quantity $\langle B_i\rangle_i$ can itself be calculated by steepest descents and is given by
$\langle B_i \rangle_i = B_i\left(\frac 12 \left(\frac{z}{\beta}-A_i\right)^{-1}\vec h_i\right) + \mathcal O (1).$
Note that the Hamiltonian Eq.~\eqref{genham} must be extensive in the number of degrees of freedom, which are the $m$ spin components. Therefore the leading term in $\langle B_i\rangle_i$ is of order $m$. The higher order cumulants can be calculated similarly. 

For simplicity, we will only consider the case $B_i(\vec s_i)=0$ and $\vec h_i$ and $A_i$ small in the following, which is valid in the vicinity of the phase transition. However, an extension to higher orders is in principle straightforward. For small $\vec h_i$ and $A_i$ Eq.~\eqref{saddle5} can be solved perturbatively. For bookkeeping purposes a factor of $\epsilon$ will be attached to $\vec h_i$ and a factor of $\epsilon^2$ to $A_i$. It will become clear below that this ansatz is consistent. A perturbative solution of Eq.~\eqref{saddle5} to second order in $\epsilon$ yields
\begin{widetext}
	\begin{align}
	Z_i&=\exp \left\{
	\frac{\beta^2J_i}{2z_i^{00}}\epsilon\vec h_i\vec s_0
	+\frac{\beta^3J_i^2}{4(z_i^{00})^2}\epsilon^2\vec {s_0}^T A_i\vec s_0
	-\frac{\beta^4J_i^2}{4m(z_i^{00})^2\sqrt{1+4\beta^2J_i^2}}\epsilon^2(\vec s_0 \vec h_i)^2+\text{const.} + \mathcal O (\epsilon^3)\right\}.
	\label{Z04}
	\end{align}
\end{widetext}
Here $z_i^{00}=\frac{1}{4}\left(1+\sqrt{1+4\beta^2J_i^2}\right)$ is the limit of $z_i^0$ as $\vec h_i\to 0$. From Eq.~\eqref{Z0} $Z_0$ can be assembled and, setting $\epsilon=1$, the new $\vec h_0$ and $A_0$ can be read off. They are given by
\begin{align}
	\vec h_{0}&=\sum_{i=1}^k\frac{\beta J_i\vec h_i}{2z_i^{00}}
	\label{h02} \\
	A_{0}&=\sum_{i=1}^k\left(\frac{\beta^2J_i^2}{4(z_i^{00})^2}A_i
	-\frac{\beta^3J_i^2}{4m(z_i^{00})^2\sqrt{1+4\beta^2J_i^2}}\vec
	h_i\otimes\vec h_i\right).
	\label{A0}
\end{align}
The symbol $\otimes$ denotes the tensor product. Eq.~\eqref{Z04} shows that the new $\vec h_0$ and $A_0$ are of order $\epsilon$ resp.\ $\epsilon^2$ such that the ansatz made above is indeed consistent.

Comparison of Eqs.~\eqref{h01} and \eqref{h02} shows that the self-consistent effective field approximation and the solution of the full problem are identical to first order in $\vec h_i$.

\section{Calculation of the phase transition}
\label{phase}
So far we have described how to iterate on the Bethe lattice with a given realization of the disorder. In order to describe the typical behaviour, we need to average over the disorder. For simplicity, we will focus on the self-consistent effective field approach first. 

\subsection{Phase transition at the self-consistent effective field level}
Instead of having a certain cavity field at each site, we now need to
find the distribution $Q(\vec h)$ of cavity fields. Under the
assumption that for $m=\infty$ the spin glass is replica symmetric, there is only one such distribution which can be found by solving the functional fixpoint equation
\begin{align}
Q(\vec h) &= E_J \int \left(\prod_{i=1}^k d^m h_i\,Q(\vec h_i)\right)\delta(\vec h - \vec h_0(\{J_i\},\{\vec h_i\})),
\label{Qh}
\end{align}
where $\vec h_0(\{J_i\},\{\vec h_i\})$ is the solution of Eq.~\eqref{h0}. This equation reflects the fact that on average, all sites are identical such that the new distribution of fields on the left hand side is the same as that of the $k$ merged spins under the integral on the right hand side. The symbol $E_J$ stands for the average over the coupling constants $J_i$. Note that Eq.~\eqref{Qh} has the trivial solution $Q(\vec h)=\delta(\vec h)$. In the paramagnetic high temperature phase the trivial solution is expected to be stable while below the transition temperature it becomes unstable. This will be explicitly verified in the following.

The phase transition takes place when, coming from high temperatures, the true local field becomes nonzero for the first time because it is linked to the Edwards-Anderson order parameter $q_{\text{EA}}=\langle \vec s\rangle^2$ via Eq.~\eqref{spinavg}. The true local field $\vec h_t$ is the field which arises when joining $k+1$ spins at a site, and its distribution is therefore $Q_t(\vec h_t) = E_J \int \left(\prod_{i=1}^{k+1} d^m h_i\,Q(\vec h_i)\right)\delta(\vec h_t - \vec h_0(\{J_i\},\{\vec h_i\}))$ where $\vec h_0$ is the solution of Eq.~\eqref{h0} but with the sum extending to $k+1$ instead of $k$. It is thus closely related to the distribution $Q(\vec h)$ of the cavity fields, and its variance is zero if and only if the variance of $Q(\vec h)$ is zero.

We analyze the variance of the distribution $Q(\vec h)$ by regarding Eq.~\eqref{Qh} as an iterative prescription. By making a Gaussian ansatz for $Q(\vec h)$ with width $\epsilon\ll 1$ on the right hand side, we calculate the width of the resulting distribution on the left hand side. If the new width $\epsilon'$ is smaller than $\epsilon$, the iteration converges to $\delta(\vec h)$, if it is larger, the trivial solution is unstable. Using the approximative Eq.~\eqref{h01}, which is valid for small $\vec h_i$, the new width is
\begin{align}
{\epsilon'}^2 &= 4\beta^2 \epsilon^2 k \int dJ_1\, P(J_1)\frac{J_1^2}{(z_1^{00})^2}.
\end{align}
Equating $\epsilon$ and $\epsilon'$, the critical temperature can easily be evaluated for the $\pm J$-model and yields
\begin{align}
T_c &= J(1-\frac 1k).
\end{align}
In the limit $k\to\infty$, this agrees with the critical temperature of the fully connected model $T_c=J$\cite{Almeida:1978b}. For the Gaussian distribution of bonds, the inverse critical temperature is given by the solution of the integral equation
\begin{align}
	1 &= 4\beta^2 \int \frac{dx}{\sqrt{2\pi}J}\, \frac{x^2e^{-x^2/2J^2}}{(1+\sqrt{1+4\beta^2x^2/k})^2}
\end{align}
which also yields $T_c=J$ in the limit $k\to\infty$.

\subsection{Phase transition for the full model}
For the full problem, a similar equation as Eq.~\eqref{Qh} can be written down for the distribution function of the matrices $A$. However, while the new field $\vec h_0$ depends only on the old fields $\vec h_i$, the matrix $A_0$ not only depends on the old matrices $A_i$ but also on the old fields, see Eq.~\eqref{A0}. Therefore the question is whether the distribution of $A$ becomes nontrivial at a higher temperature than $Q(\vec h)$. If it did, there would be another phase transition at that temperature but if it didn't, $A$ would be slaved to the $\vec h$ and aquire a nonzero variance at the same temperature as $Q(\vec h)$. Since the calculation is essentially equal to the one shown in the preceding paragraph, we will merely quote the result that the transition temperature of $A$ (in the absence of any fields) for the $\pm J$-model would be $T_c^A=\frac{J}{\sqrt[4]{k}}(1-\frac{1}{\sqrt k})$ which is lower than that of the fields for all $k$, such that there is a common phase transition at temperature $T_c$. This is of course physically sensible and agrees with the results from the fully connected model. The same is expected to hold for all higher orders if the full model is extended to include those orders.

\section{Generalized Bose-Einstein condensation}
\label{gbec}
So far we have analyzed the phase transition of the model in the thermodynamic limit $N\to\infty$ which is implicit in the cavity method. The generalized Bose-Einstein condensation, which was found for the fully connected model\cite{Hastings:2000,Aspelmeier:2004a} and in two- and three-dimensional lattices\cite{Lee:2005a}, can however only be detected for finite $N$. The reason for this is that the dimension $n_0$ of the ground state is of order $N^\mu$ with $0<\mu<1$, so the fraction $n_0/N$ vanishes as $N\to\infty$ and the situation becomes indistinguishable from a normal Bose-Einstein condensation for which $n_0=1$. We therefore proceed to analyze the ground state of our model numerically for finite $N$.

In the ground state every spin is aligned to the direction of its local field,
\begin{align}\label{Grundzustand}
H_i\vec s_i &=\sum_{j=1}^{N} J_{ij}\vec s_j.
\end{align}
Here $H_i$ is the modulus of the local field acting on the spin $\vec
s_i$. This equation can be transformed into the eigenvalue equation
$\sum_{j=1}^{N}\left(H_i\delta_{ij}-J_{ij}\right)s_j^\alpha = 0$.
This eigenvalue equation for the inverse susceptibility matrix $A_{ij}=H_i\delta_{ij}-J_{ij}$ shows that there is at least one null eigenvalue of $A$. On the other hand, it was shown that $A$ can have at most $\sqrt{2N}$ linearly independent null eigenvectors\cite{Hastings:2000}. Because the dimension of the null eigenspace corresponds to the dimension of the space spanned by the spins\cite{Aspelmeier:2004a}, the spins condense into a $n_0$-dimensional subspace of the $m$-dimensional space they live in with $0<n_0<\sqrt{2N}$. This phenomenon was named generalized Bose-Einstein-condensation. The dimension $n_0$ can be found by iterating Eq.~(\ref{Grundzustand}) until convergence, and then diagonalizing the matrix $A$. For the fully connected model it was found that $n_0\sim N^{\mu}$ with $\mu=2/5$\cite{Aspelmeier:2004a}.

\begin{figure}
\includegraphics[width=\linewidth]{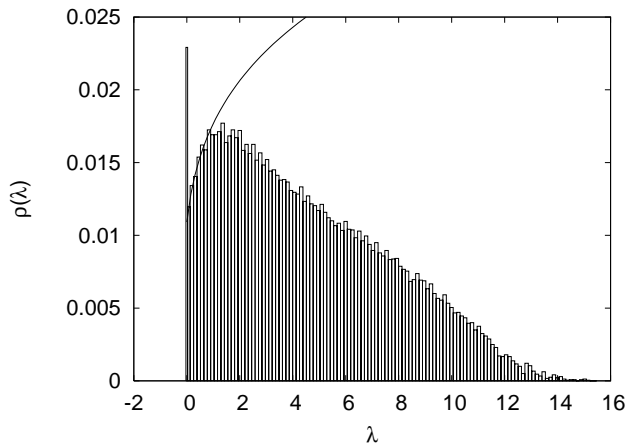}
\caption{Eigenvalue density for $N=300$ and $k+1=8$. The dashed line is proportional to $(\Delta\lambda+\lambda)^\nu$ to illustrate that the initial part of $\rho(\lambda)$ scales with the exponent $\nu$. } 
\label{SpektrumN300k8}
\end{figure}

\begin{figure}
\includegraphics[width=\linewidth]{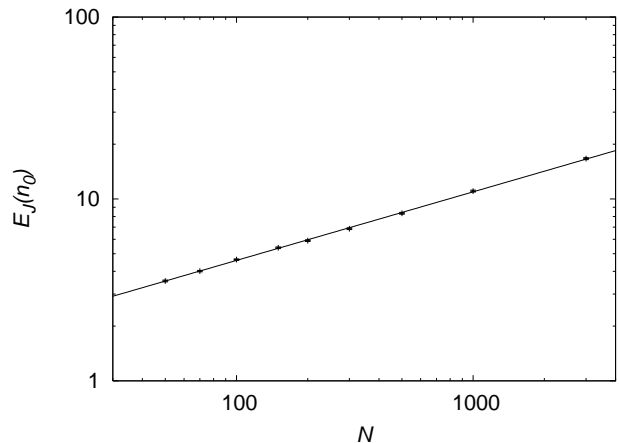}
\caption{Number of null eigenvalues for $k+1=8$ as a function of
  system size $N$ and the fit-function $f(N)\sim N^{0.377}$.}
\label{fitk8}
\end{figure}

We have repeated this computation for $m$-component spins on the Bethe lattice with $k+1=4,6,8$. The accuracy of reaching the groundstate was $10^{-7}$ for the angle between the old and the new direction of every spin $\vec s_i$. A list of system sizes and numbers of samples used can be found in Table \ref{simdata}. An example for the eigenvalue density of the matrix $A$ and the results for $E_J (n_0)$ are shown in Figures \ref{SpektrumN300k8} and
\ref{fitk8}. Table \ref{mu} shows the results for the exponent
$\mu$. The comparison with the results for the fully connected model\cite{Aspelmeier:2004a} and for $2d$- and $3d$-models\cite{Lee:2005a} puts the spin glass on the Bethe lattice in between those other models.
\begin{table}[htbp]
\caption{System sizes $N$ and corresponding numbers of samples $D$. }
\label{simdata}
\begin{center}
\begin{tabular}{c|ccccccccc}
$N$ & 50 & 70 & 100 & 150 & 200 & 300 & 500 & 1000 & 3000 \\ \hline
$D$ & 200 & 200 & 150 & 120 & 100 & 80 & 50 & 20 & 6 
\end{tabular}
\end{center}
\end{table}
\begin{table*}
\caption{Results for the exponents $\mu$, $\nu$, $x$ and $y$ and the
  ground state energy per spin and component for different
  connectivities $k+1$. The exponent $x$ was calculated from $\mu$ and $\nu$ using Eq.~\eqref{scale1}, $y$ was calculated from Eq.~\eqref{scale2}. For comparison, the known values for the fully connected, two and three dimensional models are also shown.}
\begin{center}\label{mu}
\begin{tabular}{|c|c|c|c|c|c|}\hline
 k+1 & $\mu$ & $\nu$ & $x$ & $y$ & $e_0$ \\\hline\hline
$\infty$ & 2/5 & 1/2 & 1 & 2/5 & -1 \\ \hline
8 & 0.377 ($\pm$ 0.003) & 0.234($\pm 0.0003$ ) & 1.34($\pm 0.002$)&
0.505 ($\pm$ 0.004)&-0.83($\pm 0.0016$)\\ \hline
6 & 0.362 ($\pm$ 0.004) & 0.217($\pm 0.0003$ ) & 1.45($\pm 0.002$
)& 0.525 ($\pm$ 0.006) &-0.80($\pm 0.0015$)\\ \hline     
4 & 0.352 ($\pm$ 0.005) & 0.168($\pm 0.0013$ ) & 1.57($\pm 0.0013$ )  & 0.553 ($\pm$ 0.008)  &-0.73($\pm 0.0018$)\\ \hline\hline
$d=3\cite{Lee:2005a}$ & 0.33 &--&--&--&-- \\ \hline
$d=2\cite{Lee:2005a}$ & 0.29 &--&--&--&-- \\ \hline
\end{tabular}
\end{center}
\end{table*}

\section{Scaling relations}
\label{scaling}
\begin{figure}
\includegraphics[width=\linewidth]{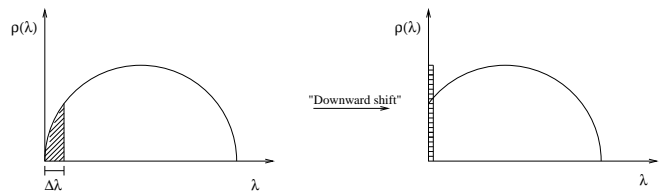}
\caption{Downward shift of the semicircular eigenvalue density of the fully
  connected $m$-component spin glass:
  $n_0$  of the smallest eigenvalues (left, shaded) become null eigenvalues
  (deltapeak, right)}
\label{Erklaerung}
\end{figure}

In Ref.\ \onlinecite{Aspelmeier:2004a} an argument was given to explain the value of $\mu=2/5$ for $n_0\sim N^{\mu}$. This value of $\mu$ arises as the result of a competition of two energy contributions to the ground state energy per spin and component $e(m,N)=\frac{E}{Nm}$. The first contribution to $e$ is the ground state energy as a function of spin components in the limit $N\to\infty$ with $m$ large but fixed. It can be deduced\cite{Bray:1981,Almeida:1978b} that this contribution is $e_0+\frac{1}{4}m^{-x}+O(m^{-2})$ with $e_0=-1$ and $x=1$. In the groundstate, $n_0$ plays the role of an effective number of spin components as the spins condense into a $n_0$-dimensional subspace, thus this contribution is $e_0+\frac{1}{4}n_0^{-x}$ for $m\ge n_0$. The second contribution consists of addititonal energy costs resulting from forcing the $N$ spins into an $n_0$-dimensional subspace. This contribution is important because we are considering the limit $m\to\infty$ \textit{before} $N\to\infty$ and $n_0$ is not fixed but proportional to $N^\mu$. The energy necessary for this can be estimated by the amount the eigenvalue spectrum $\rho(\lambda)$ of $A$ needs to be shifted in order to push $n_0$ eigenvalues to zero, see Fig.~\ref{Erklaerung}. 
Assuming that the spectrum goes as $\rho(\lambda)\sim\lambda^\nu$ for
small $\lambda$ (such a behaviour was predicted for finite dimensional systems\cite{Bray:1982}), the first $n_0$ eigenvalues occupy the space from $0$ to $\Delta\lambda\sim(\frac{n_0}{N})^{1/(\nu+1)}$, so the shift in energy will be of the same order. Combining the two terms we get
\begin{align}
e &= e_0 + c_1 n_0^{-x} + c_2 (\frac{n_0}{N})^{1/(\nu+1)}
\end{align}
with two constants $c_1$ and $c_2$. Minimizing this relation with respect to $n_0$ yields the scaling law
\begin{align}
\mu &= \frac{1}{x(\nu+1)+1}.
\label{scale1}
\end{align}
Inserting the values $x=1$ and $\nu=1/2$ for the fully connected model, we get $\mu=2/5$ as needed. The way we have presented the argument here, however, is more general and applies to the Bethe lattice and finite-dimensional lattices as well.

\begin{figure}
  \includegraphics[width=\linewidth]{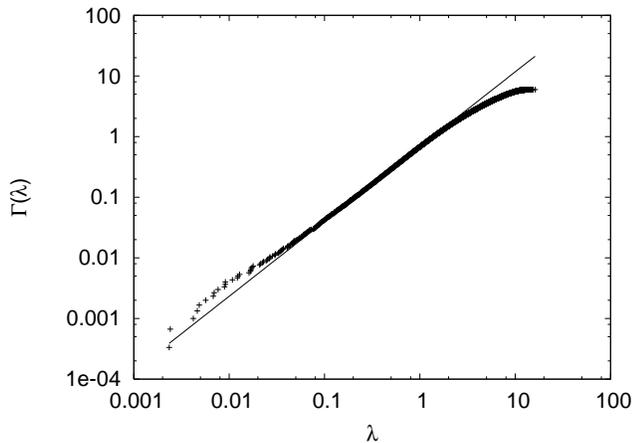}
\caption{The integrated eigenvalue density $\Gamma(\lambda)$ for $N=3000$ and
  $k+1=8$. The straight line has slope $\nu+1=1.234$. } 
\label{Int-ewDichteN3000k8}
\end{figure}

For the Bethe lattice we do not know the value of $x$ but the scaling
relation Eq.~\eqref{scale1} may be used to calculate it from $\mu$,
which can be measured as described above, and $\nu$, which may be
obtained numerically from the eigenvalue spectrum $\rho(\lambda)$. For
the latter, it must be kept in mind that we measure the spectra
\textit{after} the downward shift, i.e.\ the spectrum goes as
$\rho(\lambda)\sim (\lambda+\Delta\lambda)^\nu$ (excluding the null
eigenvalues), see Fig.~\ref{SpektrumN300k8}. The numerical fit was done
for $N=3000$ where $\Delta \lambda$ is smallest and we used the
integrated eigenvalue density $\Gamma(\lambda)$ which allows for higher precision since
there is no need for binning. An example of this fit is shown in
Fig.~\ref{Int-ewDichteN3000k8} for $k+1=8$. The results for the
exponent $\nu$ are shown in Table \ref{mu}.

\begin{figure}
\includegraphics[width=\linewidth]{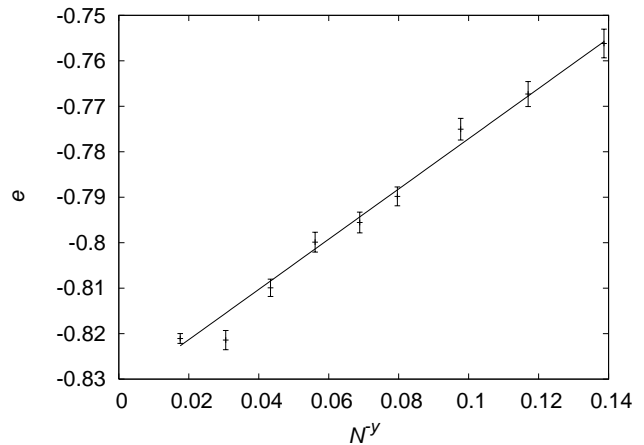}
\caption{Ground state energy per spin and component for $k+1=8$ as a function of $N^{-y}$ with $y$ calculated from the scaling relation Eq.~\eqref{scale2}. The straight line extrapolates to the ground state energy $e_0$ in the limit $N\to\infty$. } 
\label{grundzustandsenergiek8}
\end{figure}

\begin{figure}
\includegraphics[width=\linewidth]{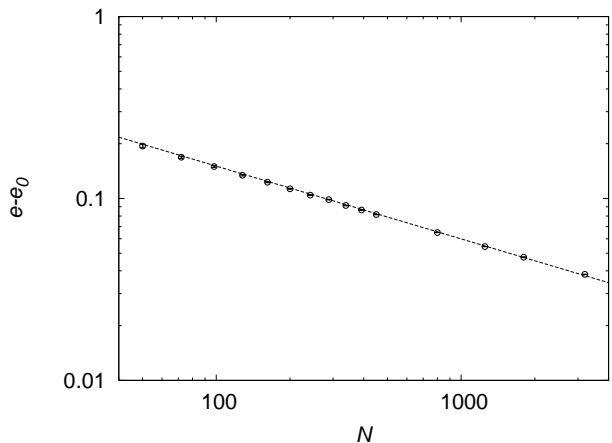}
\caption{Ground state energy per spin and component for the fully connected model. The straight line has slope $-2/5$. The error bars are smaller than the point size.}
\label{groundstateSK}
\end{figure}

In addition to Eq.~\eqref{scale1} we can obtain an entirely new scaling relation by making a scaling ansatz for the ground state energy per spin and component $e(m,N)$ in the following form,
\begin{align}
e(m,N)-e_0 &= m^{-x}F(m N^{-\mu})
\end{align}
with a scaling function $F(z)$. When we let $N\to\infty$ while keeping $m$ fixed, this yields $e-e_0\sim m^{-x}$ as required by the definition of $x$ above (provided $F(0)\ne 0$). On the other hand, when $m\ge n_0\sim N^\mu$, we know that the ground state energy becomes independent of $m$, so $F(z)=\text{const.}\times z^x$ for $z\ge n_0 N^{-\mu}$ (note that this really is an equality, not just an approximation). From this scaling ansatz it follows that $e(\infty,N)-e_0 \sim N^{-\mu x}$, i.e.\ the finite size scaling of the ground state energy with the number of spin components $m$ taken off to infinity \textit{before} $N$ scales with an exponent 
\begin{align}
	y&=\mu x = \frac{1-\mu}{1+\nu}.
	\label{scale2}
\end{align}
The second half of this equation stems from eliminating $x$ using Eq.~\ref{scale1}. In Fig.~\ref{grundzustandsenergiek8} we show that this scaling law is, within numerical precision, fulfilled for $k+1=8$.
The same goes for the other values of $k+1$ we tested (data not shown). Unfortunately, the data is too noisy to deduce the exponent $y$ directly from it. For the fully connected model, Eq.~\eqref{scale2} predicts $y=\mu=2/5$. Fig.~\ref{groundstateSK} shows that this is indeed observed.
In addition to the exponent $y$, the limiting ground state energy $e_0$ can be obtained by extrapolating the data in Fig.~\ref{grundzustandsenergiek8} to $N=\infty$. The results are listed in Table \ref{mu}. As $k$ increases, $e_0$ converges towards $-1$, the value for the fully connected model.

\section{Conclusion}
\label{conclusion}
In this work we have analyzed the $m$-component spin glass on the Bethe lattice. By extending the cavity method, we have shown that there is a phase transition to a replica symmetric state at a finite temperature for all connectivities $k+1>2$. The phase transition is a generalized Bose-Einstein condensation as for the fully connected model\cite{Aspelmeier:2004a}. We have examined four different zero temperature exponents: $\mu$, which describes the scaling of the dimension of the ground state with $N$ (for $m=\infty$), the exponent $\nu$ of the singular part of the spectrum of the inverse susceptibility matrix, $x$, the exponent which determines the scaling of the ground state energy per spin and component with $m$ (for $N=\infty$), and $y$, the scaling exponent of the same quantity with $N$ (for $m=\infty$). These exponents are connected via two scaling laws, Eqs.~\eqref{scale1} and \eqref{scale2}. As we never used the particular structure of the Bethe lattice in the derivation, we expect the scaling laws to hold in finite dimensions as well. Additionally, the exponents should be the same in the whole low temperature phase. The exponent $\nu$, for instance, should be the same at finite temperature because it was argued in \cite{Aspelmeier:2004a} that the eigenvalue spectrum of the inverse susceptibility matrix becomes stuck as soon as the temperature goes below the critical temperature. This is also what happens in the spherical model\cite{Almeida:1978b}. At any rate, the scaling laws may be used to obtain the exponent $x$, which is a quantity of interest because it extrapolates towards finite $m$ for a system in the thermodynamic limit $N=\infty$, from the exponents $\mu$, $y$ and $\nu$ which are defined for $m=\infty$, a case which is often computationally and analytically easier to handle.

Our exponent $y$ is analogous to the shift exponent $1-\Theta_s/d$ as discussed in Ref.\ \onlinecite{Bouchaud:2003} for Ising spin glasses. There it was found that $1-\Theta_s/d=2/3$ for all mean-field models considered including the Bethe lattice with connectivities $3,4,6$ and $10$. For the fully connected model, this was also confirmed by Billoire\cite{Billoire:2006}. This is clearly not the case for the $m=\infty$ model, see Table \ref{mu}. For the fully connected model, we found $y=2/5$ and increasing values for smaller connectivities. The origin of this difference in behaviour is not clear to us.

In this work we have presented the leading contribution as $m\to\infty$. It should be pointed out, however, that our modification of the cavity method can easily be extended to obtain higher order corrections in $1/m$. Work along these lines is in progress.

\section*{Acknowledgements}
We would like to thank M. A. Moore for many stimulating discussions. TA acknowledges support by the Deutsche Forschungsgemeinschaft (grant Zi-209/7).
AB acknowledges support by the {\it VolkswagenStiftung} (Germany)
within the program ``Nachwuchsgruppen an Universit\"aten''.
\bibliography{LiteraturDB}
\end{document}